\documentclass[aps,prb,twocolumn,groupedaddress]{revtex4}

\usepackage{graphicx}

\begin{document}


\title{Energetics of positron states trapped at vacancies in solids}


\author{I. Makkonen}
\email[Electronic address: ]{ima@fyslab.hut.fi}
\affiliation{Laboratory of Physics, Helsinki University of Technology, P.O. Box 1100, FI-02015 HUT, Finland}
\author{M. J. Puska}
\affiliation{Laboratory of Physics, Helsinki University of Technology, P.O. Box 1100, FI-02015 HUT, Finland}


\date{\today}

\begin{abstract}

We report a computational first-principles study of positron trapping 
at vacancy defects in metals and semiconductors. The main emphasis is 
on the energetics of the trapping process including the interplay between 
the positron state and the defect's ionic structure and on the ensuing 
annihilation characteristics of the trapped state. For vacancies 
in covalent semiconductors the ion relaxation is a crucial part of 
the positron trapping 
process enabling the localization of the positron state. However, 
positron trapping does not strongly affect the characteristic features 
of the electronic structure, e.g., the ionization levels change only
moderately. Also in the case of metal vacancies the positron-induced ion 
relaxation has a noticeable effect on the 
calculated positron lifetime and momentum distribution of annihilating
electron-positron pairs.  

\end{abstract}


\maketitle



%



\section{Introduction}

Positron annihilation spectroscopy~\cite{Krause-Rehberg99} is widely
used in studying vacancy-type defects in metals and semiconductors. 
This is based on the fact that vacancies, in neutral or negative
charge states, act as efficient positron traps due to the reduced
repulsion of positive ions. Also the electron density is reduced at
vacancies and by measuring the ensuing lifetime increase with respect
to the delocalized 
positron bulk state one can estimate the extent of the open volume 
in defects. The momentum distribution of annihilating electron-positron 
pairs measured by Doppler broadening spectroscopy 
is at high momenta specific for the annihilating core electrons
and thereby it enables the chemical identification of atoms surrounding 
vacancies. 

For a given sample, the identification of the most abundant
open-volume defect type, such as vacancy and vacancy agglomerates or
vacancy-impurity complex, is based on the knowledge of the general behavior 
of positron annihilation results, i.e., on the information how the
measured annihilation characteristics depend on the open volume or the
chemical environment of defects. This knowledge has been acquired
by measuring well-characterized reference samples including 
well-annealed (perfect bulk) materials as well as defected materials 
such as those containing monovacancies due to electron irradiation. 
Moreover, theoretical predictions of positron annihilation 
characteristics significantly support the defect
identification.~\cite{Puska94}

In order to interpret experimental results it is also important 
to understand the positron trapping process in detail. The trapping 
coefficient is an important quantity in determining defect 
concentrations. Its values have been estimated also theoretically
for model systems by assuming that the positron gives in the trapping 
process its binding energy to excited electron-hole pairs or to
phonons.~\cite{Hodges70,Nieminen79,Puska90}
In the present computational work our main theme is the effect
of the positron on the ion lattice during the trapping process.
Because the ion vibration frequency, which is of the order of the Debye
phonon frequency, is much larger than the positron annihilation
rate the ions around the defect have time prior to positron 
annihilation to relax to minimize the total energy of the
defect-positron system. The ion relaxation affects the localization
of the positron state and the annihilation characteristics. Below
we argue using first-principles total-energy calculations that
the positron-induced lattice relaxation is indispensable for
the existence of localized positron states at vacancies in covalent 
semiconductors and thereby it completely determines the ensuing
annihilation characteristics. In these systems, the strong influence
of the positron is possible because the energy landscape as a function 
of the ion positions around the vacancy is very flat. Actually, the
effect of the trapped positron is found so strong that it practically cancels
the possible symmetry-lowering Jahn-Teller distortion of the vacancy.  
For vacancies in metals the influence of the trapped positron on the
ion positions and especially on the energetics of the trapping
process is smaller. However, the trapped positron causes
a small increase in the open vacancy volume and thereby noticeable
chances in the positron lifetime and in the momentum distribution of
the annihilating pairs. It is important to note this from the modeling
point of view.

In the case of semiconductors, positron annihilation has been used 
also to extract
detailed information about the electronic structures of the
defects, i.e., to determine the ionization levels of vacancy-type
defects~\cite{Corbel88,Saarinen91a,Saarinen91b,LeBerre95,Kuisma97,Kauppinen99,Arpiainen02,Tuomisto05}
or
just to probe the changes in their charge states.~\cite{Makinen92} 
In these experiments
the charge state of the defect changes due to the thermal 
ionization or by illumination with light. The charge state change is 
then seen as a change in the positron lifetime reflecting 
electronic-structure-induced changes in the ion positions or as 
a dramatic change in the positron trapping rate when positive 
defects do not trap positrons. The question rising immediately in the
first case is how much the positron-induced ionic relaxation affects
the positions of the ionization levels by modifying or eventually
breaking bonds between ions next to the vacancy. Our prediction is that
although the changes in the ionic structure are rather large their
effect on the ionization levels is minor. 

Finally, we would like to point out that the comparison of calculated
positron annihilation characteristics with the measured ones constitutes
the fundamental test for theories describing electronic properties
of materials and the positron-electron interactions as well as for
many computational approximations. For delocalized positron states in
perfect bulk solids there exist several systematic 
comparisons~\cite{Puska83,Puska89,Barbiellini96,Makkonen06} but for
positron states trapped at vacancy defects comparisons treating
several materials and systems on the same
footing are scarce. The reason may be in difficulties arising in the 
theoretical description, e.g.\ in the density-functional 
theory~\cite{Hohenberg64,Kohn65} (DFT) the local-density approximation 
(LDA) for the electron exchange and correlation  underestimates the 
energy band gap in 
semiconductors which may have severe consequences on the localized
electron states and the ionic structure at defects. Moreover, the 
electron-positron correlation effects are known worse for localized 
positron states than for delocalized ones. Lastly, the broken 
translational symmetry leads to computational approximations such as the
supercell method which requires large computer resources in order
to show convergence of the results with respect to the supercell size.
The aim of the present study is to remedy the situation by providing
results for a representative set of materials. We consider metals with
different lattice structures (close-packed Al, Cu, Mg, and 
body-centered-cubic Fe) and elemental (Si, Ge) and compound
semiconductors (GaAs, GaN) with different degrees of bond ionicity.
The structure of the present paper is as follows. In Sec.~\ref{theory}
we review shortly the theory and computational methods used. The results
are given and described in Sec.~\ref{results} and they are discussed
along comparisons with experimental results in
Sec.~\ref{discussion}. Section~\ref{conclusions} presents our
conclusions.

\section{Theory and computational details}\label{theory}

\subsection{Theoretical models}\label{models}

We perform first-principles electronic-structure calculations based on
DFT for various
vacancy defects in metals and in semiconductors. These calculations
give the ionic positions by requiring that the total energy is minimized.
This is equivalent to vanishing of the Hellman-Feynman forces on ions, 
calculated from the ground-state electron density.
The trapped positron 
state at a defect can be included by generalizing to the two-component 
density-functional theory~\cite{Boronski86} (2CDFT). For defects in 
semiconductors calculations optimizing the electronic and ionic 
structures as well as the positron density within the 2CDFT have 
appeared.~\cite{Gilgien94,Puska95,Saito96,Tang97,Makhov05} 

In the present work we apply the so-called conventional scheme
in which $(i)$ the localized positron density does not directly affect
the average DFT electron density (the positron and its screening 
electron cloud form a neutral quasiparticle entering the system) 
and $(ii)$  the positron state and annihilation characteristics are 
calculated in the LDA and at the zero-positron-density 
limit of the electron-positron correlation functionals.
For example, this means that the potential entering the 
single-particle
equation for the positron state
$\psi_+(\mathbf{r})$ reads as
\begin{equation}\label{pospot}
V_{+}(\mathbf{r})=\phi(\mathbf{r})+V_{\text{corr}}\textbf{(}n_{-}(\mathbf{r})\textbf{)},
\end{equation}
where $\phi (\mathbf{r})$ is the Coulomb potential due to electrons
and nuclei, $n_{-}(\mathbf{r})$ the electron density and
$V_{\text{corr}}(n_{-})$ is the electron-positron correlation energy~\cite{Boronski86}
for a positron in a homogeneous electron gas with density $n_{-}$.
It has been shown that the effects of the above two approximations
largely cancel each other's effects so that the conventional scheme and
2CDFT results for positron annihilation characteristics, for
the total energy of the defect-positron system and for the positron
trapping energy are very 
similar.~\cite{Boronski86,Puska95} 
Besides due to the conceptual simplification, we prefer the conventional 
scheme also because the 2CDFT functionals for finite positron densities are 
not accurately known.

When we relax the ions surrounding a defect with a trapped positron
we minimize the total energy which in the conventional scheme is 
the sum of the DFT total energy for the electron-ion system and
the positron energy eigenvalue. Thus,
although our calculation is not a self-consistent 2CDFT calculation the
positron state and the 
electron density are coupled via
the ionic structure. In practice, we calculate the positron-induced
Hellman-Feynman forces on ions using the so-called atomic superposition 
method (for details, see Ref.~\onlinecite{Makkonen06}).

The total annihilation rate $\lambda$ which is the inverse of the
positron lifetime $\tau$ is obtained from the overlap integral
of the electron density $n_-({\bf r})$ and positron density 
$n_+({\bf r})=\vert \psi_{+}({\bf r})\vert^2$
\begin{equation}
\lambda = \frac{1}{\tau}=\pi r_e^2 c \int n_-({\bf r})n_+({\bf r})
\gamma \textbf{(}n_-({\bf r})\textbf{)} d{\bf r}.
\end{equation}
Above, $\gamma(n_-)$ is the enhancement factor for a positron
a homogeneous electron gas~\cite{Boronski86} with density $n_{-}$,
and $r_e$ and $c$ are the classical
electron radius and the speed of light, respectively. We calculate
the momentum distribution $\rho({\bf p})$ of the annihilating 
electron-positron pairs using the state-dependent 
enhancement scheme.~\cite{Alatalo96,Barbiellini97} I.e.,
\begin{equation}
\rho({\bf p})=\pi r_e^2 c \sum_j \gamma_j \bigg\vert \int e^{-i{\bf p} \cdot {\bf r}}\psi_+({\bf r})
\psi_j({\bf r}) d{\bf r}\bigg\vert ^2,
\end{equation}
where $\psi_j(\mathbf{r})$ and $\gamma_j$ are the wave
function and the state-dependent enhancement factor (in the LDA) for
the electron state $j$.

Besides the agreement with the 2CDFT the results obtained with the 
conventional scheme compare reasonably well also with 
experiment.~\cite{Puska95,Makkonen06} First and foremost,
the measured changes in the positron lifetime and in the relative
changes in the core annihilation rate between the vacancy and
bulk states are reproduced.

The most important aspect of the present work is the energetics
of the defect-positron system. We define the positron trapping 
energy at a vacancy defect (the energy released in the trapping process)
as the total energy difference between the systems of $(i)$ a defect
and a delocalized positron and $(ii)$ the same defect trapping a 
positron. Within the conventional scheme we obtain 
\begin{eqnarray}
E_{t} & = & \Delta
E_{\text{tot}}=(E+\varepsilon^{+}_{\text{bulk}})-
(E_{e^{+}}+\varepsilon^{+}_{\text{defect}}) \nonumber\\
& = & (\varepsilon^{+}_{\text{bulk}}-\varepsilon^{+}_{\text{defect}})
-(E_{e^{+}}-E),
\label{trappingE}
\end{eqnarray}
where $E$ is the total energy of the electron-ion system of the defect
supercell without the localized positron, $E_{e^{+}}$ that with the
localized positron. $\varepsilon^{+}_{\text{bulk}}$ and
$\varepsilon^{+}_{\text{defect}}$ are the energy eigenvalues of the
positron in the delocalized bulk state and in the localized state at the
vacancy, respectively. The last form in Eq.~(\ref{trappingE}) shows that
the trapping energy consists of the decrease of the positron energy
eigenvalue and the increase in the (strain) energy stored in the ion
lattice. In general, the Kohn-Sham eigenvalues in the
2CDFT have no physical meaning but as we have only one positron in the
lattice and we use the conventional scheme the
electron-positron
interactions  affect only the positron energy eigenvalue
$\varepsilon^{+}$ and the above analysis is justified.

The ionization level $\varepsilon(Q/Q')$ between the charge states
$Q$ and $Q'$ of a defect is defined as the position of the 
chemical potential $\mu_{e}$ relative to the top of the valence band $E_{v}$
so that the total energies of these two charge states are equal. I.e.,
we solve for $\mu_{e}$ in
\begin{equation}
E_{\text{tot}}^{Q}+Q(E_{v}+\mu_{e})=E^{Q'}_{\text{tot}}+Q'(E_{v}+\mu_{e}),
\label{ionlevel}
\end{equation}
where $E^{Q}_{\text{tot}}$ is the total energy of the supercell with
the defect in the charge state $Q$ and the term $Q(E_{v}+\mu_{e})$
arises because $Q$ electrons are added to ($Q > 0$) or taken from
($Q < 0$) the electron reservoir at the chemical potential level 
$E_{v}+\mu_{e}$. 

\subsection{Calculation methods}

In the supercell approach we use one has to take care of well-known
artifacts. First, the energy eigenvalue
$\varepsilon^{+}_{\text{bulk}}$ of the delocalized positron in
Eq.~(\ref{trappingE}) and
the valence band maximum $E_{v}$ in Eq.~(\ref{ionlevel})
are taken from the calculation for the perfect periodic bulk material.
Because the energy zeroes differ between different supercells we align
the effective potentials for the defect and bulk supercells
far from the defect both in the case of electrons and 
the positron. Second, in order to avoid long-range Coulomb interactions 
between charged supercells in the superlattice we use a neutralizing 
uniform background charge. The unphysical energy terms due to 
the monopole-monopole interactions between the periodic images of 
the defects are corrected by the method by Makov and
Payne.~\cite{Makov95} However, in
the case of defects in GaN we do not apply these corrections
because they lead to an overcorrection. Since we are
primarily interested not in the absolute values of the ionization
levels but in their changes due to the localization of the positron
the energy corrections are actually not of utmost importance.
 
Our computational methods are described in more detail in 
Ref.~\onlinecite{Makkonen06} and here we will give only the 
main features. We perform electronic structure calculations
within the LDA (Ref.~\onlinecite{Perdew81}).
The description of the
electron-ion interaction is based on the projector augmented-wave (PAW)
method~\cite{PAW} implemented in the plane-wave code 
\textsc{vasp} (Refs.~\onlinecite{Kresse96a,Kresse96b,PAWKresse}).
Using the PAW total charge density including the free atom core 
electrons the positron potential is constructed according to 
Eq.~(\ref{pospot}) 
and the lowest-energy positron state
is calculated on a three-dimensional real-space point
grid. 
The lattice constants of the perfect bulk lattices
are optimized 
and used in the defect calculations to
define the supercell volume. For Si, Ge, and GaAs we use cubic 216-atom
supercells. The Brillouin zone is sampled in the case of Si and GaAs 
using a $2^{3}$ Monkhorst-Pack (MP) $\mathbf{k}$ point 
meshes~\cite{Monkhorst76} whereas for Ge we use the $L$-point sampling
in order to avoid the artificial hybridization of the deep level and
band states in the LDA (see Ref.~\onlinecite{Coutinho06}). Cutoff energies 
for Si, Ge, and GaAs are 246~eV, 270~eV, and 209~eV, respectively. 
We model wurtzite GaN using an orthorhombic 96-atom supercell, a $3^{3}$ MP
$\mathbf{k}$-mesh and a cutoff energy of 400~eV. The 
hexagonal close-packed (hcp) Mg is modeled using an orthorhombic 48-atom
supercell, a $8\times6\times 6$ $\mathbf{k}$-mesh and the cutoff of
263~eV. For the body-centered cubic (bcc) Fe we calculate the magnetic ground
states using a cubic 54-atom supercell,
a $8^{3}$ MP $\mathbf{k}$-mesh and a cutoff energy of 268~eV. For
defects in the face-centered cubic (fcc) metals Cu and Al we use
cubic 108-atom supercells, a $6^{3}$ and $8^{3}$ MP
$\mathbf{k}$-meshes and the cutoff energies of 342~eV and 301~eV, respectively.

\section{Results}\label{results}

\subsection{Ion relaxation in positron trapping at
  vacancies}\label{relaxations}
 
We study first the energetics of the trapping process, i.e., the
interplay between the lowering of the positron energy eigenvalue
and the energy stored in the strained lattice around the vacancy.
Vacancies in Al and in Si represent two very different types of
behaviors. In order to facilitate the scanning of the energy
landscape when the ions relax due to the positron-induced forces
we consider only the breathing-type ion relaxation. Thus, 
the point symmetries of the Al and Si vacancies are constrained
to be $O_h$ and $T_{d}$, respectively. The reaction coordinate
is the relaxation of the nearest-neighbor ions of the vacancy
from their ideal lattice positions. The positions of the
other ions in the supercell are optimized. 

\begin{figure}
\includegraphics[width=0.8\columnwidth]{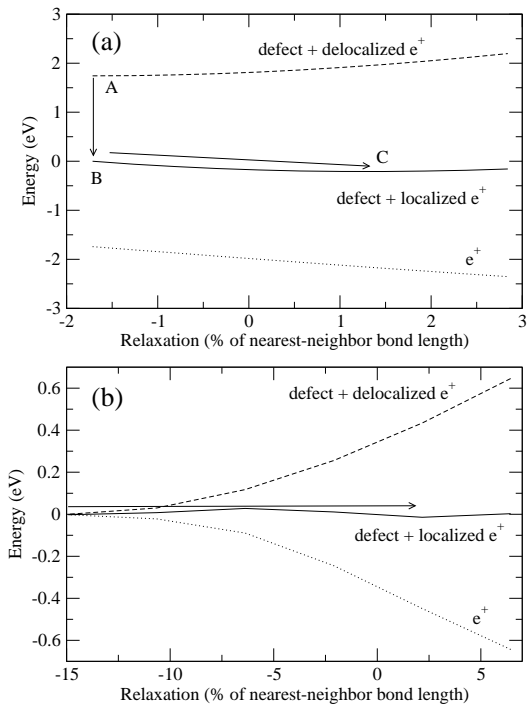}%
\caption{Configuration-coordinate diagrams for (a) the monovacancy in
  Al and (b) the neutral monovacancy in Si. The dotted lines show the
  positron energy eigenvalue (relative to the one in perfect bulk), dash
  lines correspond to the energy of the lattice and the delocalized
  positron and solid lines the total energy of
  the defect-positron system as functions of the relaxation of the
  vacancy. Positive (negative) sign denotes outward (inward)
  relaxation. The points A, B, and C denote different stages in the
  positron trapping process.\label{confcoord}}
\end{figure}

The results for the Al vacancy are shown in Fig.~\ref{confcoord}(a).
The energy of the electron-ion system (the uppermost curve) and the 
positron energy eigenvalue (the lowest curve) as well as their sum, 
the total energy of the system (the curve in the middle), are shown as 
a function of the relaxation of the nearest-neighbor ions.
The energy zero is chosen to be the total energy of the vacancy
and the trapped positron with ions relaxed without positron-induced
forces. Then the uppermost curve corresponds also
to the total energy of the vacancy and a delocalized positron. 
The smallest relaxation shown corresponds to the equilibrium ion 
positions of the Al vacancy without the trapped positron. 
The picture of the positron trapping process is clear. First, a fast 
(fast compared to the time scale of ionic movement) vertical
Franck-Condon shift of the
positron from the delocalized bulk state to the localized ground state
at the vacancy takes place via electron-hole excitation
[A$\rightarrow$B in Fig.~\ref{confcoord}(a)]. Then,
the ions move slightly outward to minimize the total energy of the
vacancy-positron system (B$\rightarrow$C). The escape of the
positron from the trapped state via thermal processes, the so-called 
detrapping process, is very unlike due to large separation of two
uppermost total energy curves.

The energetics for the neutral Si vacancy trapping a positron
is depicted in Fig.~\ref{confcoord}(b). Actually, in this
case we cannot find a bound positron state for the strongest
inward relaxations and therefore the curves join at the leftmost
point corresponding to the $T_d$ symmetric equilibrium relaxation
without the trapped positron. At smaller inward relaxations
a bound state exists and the positron energy eigenvalue decreases.
Surprisingly, the energy stored in the electron-ion
system and the lowering of the positron energy eigenvalue 
cancel each other rather accurately and the sum curve is
very flat over a large range of ionic relaxation. The total 
energy shows a maximum and at small outward relaxations a
minimum which gives a trapping energy [Eq.~(\ref{trappingE})] 
of 0.05 eV. In the trapping process a
thermalized positron would ``clear'' a larger empty volume and a
slightly deeper potential well for itself. In this way the
trapping process is analogous to the self-trapping of
electrons or holes in small polaron states in ionic crystals (see for
example Ref.~\onlinecite{Mott77}).
The flat total energy surface means that at
finite temperatures the ions can be quite far from their absolute
minimum energy configuration. However, the entropy contribution
to the free energy would favor a larger open volume; The situation that
all the four nearest-neighbor atoms are very close each
other (strong inward relaxation) is very unlikely due to the
small corresponding phase-space volume. As a 
consequence (assuming that positron detrapping is a vertical Franck-Condon
process) there would also be a finite effective detrapping energy
related to the distance between the two uppermost total
energy curves in Fig.~\ref{confcoord}(b). The detrapping energy would
be of the order of tenths of an eV.

In conclusion, the main characteristic differences between positron
trapping at Al and Si vacancies are the much larger change in the 
positron energy eigenvalue for the Al vacancy and the much larger
ion relaxation at the Si vacancy.  
The repulsive effect of the localized positron is stronger than one
might expect on the basis of its small charge. The zero-point motion
of the positron increases the force on the neighboring ions because
the positron density penetrates closer to their nuclei. For example, we
estimate that in the case of the Si vacancy the force due to a
classical positive unit point charge at the center of the vacancy is
only $\sim 50$\% of the force due to the localized positron.
For the Al vacancy the $O_h$ symmetry persists also without
constraints but for the neutral Si vacancy a symmetry lowering
Jahn-Teller distortion is expected.
In fact, when the defect is relaxed without a localized positron we
find a Jahn-Teller distortion with the $D_{2d}$ symmetry that lowers
our calculated trapping
energy slightly so that it even becomes negative. When the positron is
trapped at the vacancy its repulsion practically restores the $T_{d}$
symmetry of the vacancy. Within the numerical 
accuracy we can consider the trapping energy to be practically
zero. The flatness of the energy landscape will be a general
characteristic feature of the Si vacancy trapping a positron.

\subsection{Trapped positron states and annihilation at vacancies}

\begin{figure}
\includegraphics[width=1.0\columnwidth]{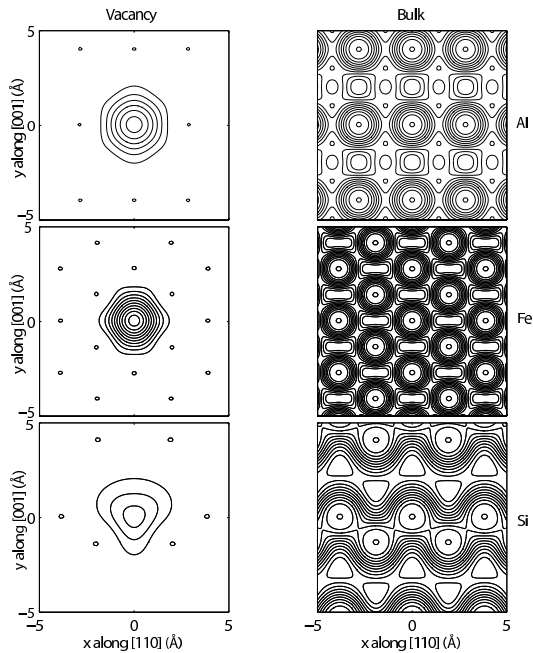}%
\caption{Positron densities in monovacancies (left, contour spacing
  0.01~\AA$^{-3}$) in Al, Fe and Si and corresponding perfect lattices
  (right, contour spacing one tenth of the maximum value). The dots in
  the figures denote the locations of the nuclei on the
  plane.\label{positrondensities}}
\end{figure}

Next we present our first-principles results for various semiconductors
and metals. Figure~\ref{positrondensities} shows the calculated positron 
densities at Al, Fe, and Si vacancies and in corresponding defect-free
lattices. Thus, examples of fcc and bcc metals and 
tetrahedrally-bonded semiconductors are considered.
The vacancies in metals localize the positron state effectively whereas in Si
the positron density tends to leak along the open interstitial
channels, which is reflected also in the smaller maximum value of the
positron density. Because of the higher coordination number the Coulomb
repulsion due to the nuclei is larger
in the interstitial regions in the fcc and bcc lattices than in the
open interstitial channels in the diamond structure of Si. Therefore
the positron energy eigenvalue will decrease in the trapping process
more in the fcc and bcc metals than in semiconductors

More quantitatively, we calculate positron trapping energies at 
vacancies and analyze how localized positrons affect the volumes 
and symmetries of the defects. Moreover, for vacancies in 
semiconductors we determine thermodynamical ionization levels in 
the energy band gap with and without a trapped positron.
Our main results are presented in Table~\ref{energetics} and
they are discussed in the following subsections.

\subsubsection{Trapping energies}

The positron trapping energies with components giving 
the lowering of the positron energy eigenvalue and the energy
stored in the ion lattice are shown on the leftmost data columns 
in Table~\ref{energetics} [see the last form in  Eq.~(\ref{trappingE})].
Several trends can be seen. First of all, the positron trapping
energy at vacancies ($V$) in metals is typically clearly larger than that in
semiconductors, i.e., the values for $V_{\text{Al}}$, $V_{\text{Fe}}$,
and $V_{\text{Cu}}$
are of the order of 2~eV whereas the values for $V_{\text{Si}}$
and $V_{\text{Ga}}$ in GaAs are at most a few tenths
of an eV. As can be seen in Table~\ref{energetics} this difference
originates mainly from the fact that the lowering of the positron
energy eigenvalue
is larger in metals than in semiconductors. This in turn reflects
the reduction of the positron-nucleus Coulomb repulsion which is
larger when a vacancy is created in metal lattices with a larger
atomic density and higher coordination number than in tetrahedrally 
coordinated semiconductor 
lattices with large open interstitial channels. However, there are
exceptions from this general trend. Among the 
vacancies in metals, $V_{\text{Mg}}$ has a very low positron trapping energy
which reflects the relatively low atom and electron densities.
$V_{\text{Ga}}$ in GaN has a positron trapping energy similar to
metals. This is no wonder because due to the size difference
between the Ga and N atoms the Ga atom density in GaN is more
than 60\% of that in the Ga metal. The high atom density increases 
the positron-nucleus repulsion in the perfect GaN lattice and the 
lowering of the energy eigenvalue in trapping.

%
\begin{table*}
\caption{Positron trapping energies and their decompositions into the
  decrease of the positron energy eigenvalue and increase of the
  energy of the lattice, ionization
  levels, relative volume changes and resulting point symmetry groups,
  positron lifetimes $\tau$ and relative $W$ parameters for various
  vacancy defects in
  different charge states $Q$ in bulk solids. Results are
  calculated with a positron trapped at the defect except for the ones
  in parenthesis which are obtained without the trapped
  positron. Ionization levels are given with respect to the valence 
  band maximum.  The relative changes $(V-V_0)/V_0$ in the vacancy open 
  volume are calculated from the volumes of the polyhedra defined by 
  the nearest-neighbor atoms in the ideal ($V_0$) and relaxed ($V$) lattice
  positions. Negative (positive) values correspond to inward (outward)
  relaxation of the nearest-neighbor atoms. The positron lifetimes
  at defects should be contrasted with our corresponding computational
  lifetimes for the bulk solids which are 
  Si: 208, Ge: 213, GaAs: 212, GaN: 131, Fe: 87, Mg: 219, Cu: 95, 
  and Al: 159~ps.
\label{energetics}}
\begin{ruledtabular}
\begin{tabular}{lrcccrrccc}
Defect & $Q$ & $E_{t}$&
$\varepsilon^{+}_{\text{bulk}}-\varepsilon^{+}_{\text{defect}}$ &
$E_{e^{+}}-E$ & $\varepsilon$($Q/Q$--1) & $(V-V_{0})/V_{0}$ & Symmetry
& $\tau$ & $W_{\text{rel}}$\\
    &   & (eV) & (eV) & (eV) & (eV) & (\%) &  & (ps) & \\
\hline
$V$ in Si & 0 & --0.17 & 0.69 & 0.86 & 0.47 (0.47) & +20.5 (--43.1) &
$T_{d}$ ($D_{2d}$)& 260 & 0.55\\
        & --1 & --0.17 & 0.80 & 0.97 & & +13.3 (--54.8)\footnote{Split
  vacancy.}& $T_{d}$ ($D_{3d}$)\footnotemark[1] & 255 & 0.57\\
$V_{\text{Ge}}$ in Ge & 0 & & \multicolumn{2}{l}{(no minimum with
  bound e$^{+}$)} & (0.05) & (--42.5) & ($D_{2d}$) & \\
& --1 & & \multicolumn{2}{l}{(no minimum with
  bound e$^{+}$)} & & (--45.3) & ($D_{2}$) & \\
\hline
$V_{\text{Ga}}$ in GaAs & --2 & 0.39 & 0.75 & 0.36 & 0.74 (0.81) &
--13.0 (--37.3) & $T_{d}$ ($T_{d}$)& 237 & 0.68\\
& --3 & 0.46 & 0.84 & 0.38 & & --16.2 (--37.5) & $T_{d}$ ($T_{d}$) &
234 & 0.69\\
$V_{\text{As}}$ in GaAs & 0 &  & \multicolumn{2}{l}{(no minimum with bound e$^{+}$)} & (0.21) & (--41.1)  & ($D_{2d}$)\\
& --1 &  & \multicolumn{3}{l}{(no minimum with bound e$^{+}$)} & (--51.1) & ($D_{2d}$)\\
\hline
$V_{\text{Ga}}$ in GaN & --2 & 1.82 & 2.05 & 0.23 & 1.22 (1.43) &
+59.0 (+29.7)& $C_{3v}$ ($C_{3v}$) & 216 & 0.57\\
& --3 & 2.04 & 2.28 & 0.24 & & +60.7 (+29.4) & $C_{3v}$ ($C_{3v}$) &
216 & 0.55\\
$V_{\text{N}}$ in GaN & 0 & & \multicolumn{2}{l}{(no bound e$^{+}$
  state)} & (2.62) & (--7.7) & ($C_{3v}$)\\
& --1 & & \multicolumn{2}{l}{(no bound e$^{+}$
  state)} & & (--19.5) & ($C_{3v}$)\\
\hline
$V$ in bcc Fe & & 1.67 & 2.34 & 0.67 & & +8.4 (--6.2) & $O_{h}$
($O_{h}$) & 159 & 0.75\\
\hline
$V$ in hcp Mg & & 0.34 & 0.49 & 0.15 & & +6.3 (--2.9) & $C_{3h}$
($C_{3h}$) & 289 & 0.56\\
$V$ in fcc Cu & & 2.20 & 2.58 & 0.38 & & +7.4 (--3.8) & $O_{h}$
($O_{h}$) & 163 & 0.74\\
$V$ in fcc Al & & 1.89 & 2.35 & 0.46 & & +8.8 (--5.1) & $O_{h}$
($O_{h}$) & 242 & 0.78\\
\end{tabular}
\end{ruledtabular}
\end{table*}

According to Table~\ref{energetics} the energy stored in to the 
lattice relaxation during the positron trapping process is 
of the same order of magnitude, $\sim 0.5$~eV. However, the crucial
difference between typical metals and semiconductors is that for metals this
energy is only a fraction of the energy which the positron 
gains in the lowering of the energy eigenvalue 
while for semiconductors it is of the same order
of magnitude. Indeed, in the case of $V_{\text{Si}}$, 
and $V_{\text{Ga}}$ in GaAs the two components of the trapping energy
nearly cancel each other.

Our calculations predict that the trapped positron state and the
accompanying ionic relaxation at 
$V_{\text{Si}}$ is only a metastable configuration and that the global
energy minimum corresponds to an unperturbed vacancy and a delocalized
positron. The energy barrier between these two minima is only some
tenths of meV larger than the absolute value of the predicted
negative trapping energy. 
The actual situation differs from that 
depicted in Fig.~\ref{confcoord}(b) because allowing the 
symmetry-breaking the Jahn-Teller effect lowers the energy of the
vacancy relaxed without the trapped positron.

In the case of $V_{\text{Ge}}$ and $V_{\text{As}}$ in GaAs we find 
a bound positron state
when the ions are frozen at ideal lattice positions or the nearest neighbor
ions are relaxed outwards. But when we start optimizing the
ion positions the vacancy relaxes strongly inward destroying the
bound state. $V_{\text{Ge}}$ can be contrasted with $V_{\text{Si}}$;
The increase in the lattice constant from Si to 
Ge does not compensate the increase in the ion size and not even a
metastable configuration with a bound positron state is found. For 
$V_{\text{N}}$ in GaN we did not found a bound positron state even
when the ions neighboring the vacancy were frozen at positions
corresponding to reasonable  outward relaxations. In the latter case
this can be explained by the small size of the N ion. 
 
\subsubsection{Defect geometries}

The relative changes $(V-V_0)/V_0$ in the vacancy open volume are
also given in Table \ref{energetics}. Here $V_0$ and $V$ refer
to the volumes of the ideal (atoms at the ideal lattice sites) and
relaxed vacancies,
respectively. $V_0$ and $V$ are calculated as the volumes of
polyhedra restricted by the nearest-neighbor atoms of the vacancy.
The numbers in the parenthesis show that without trapped positrons 
vacancies in metals and in typical semiconductors have a tendency 
to shrink, i.e., the nearest neighbor atoms relax inward toward the 
center of the vacancy. The relaxation is large for semiconductors
whereas in metals the ions remain close to their ideal lattice
positions. In GaN the N atoms neighboring  $V_{\text{Ga}}$ relax outward
which reflects again the role of large Ga atoms in determining
the lattice constant. The trapped positron 
increases the open volume. The effect is
very dramatic in the case of semiconductors in which the
volume increase of the vacancies is several tens of percents. For the neutral
$V_{\text{Si}}$ it is even of the order of 60\%. The changes in the vacancy
open volume are reflected in the positron lifetime and the momentum
density of the annihilating electron-positron pairs (see
Ref.~\onlinecite{Makkonen06}). The effect is strong also in the case of
metals although the changes in the ionic relaxations are smaller.

\begin{figure}
\includegraphics[width=\columnwidth]{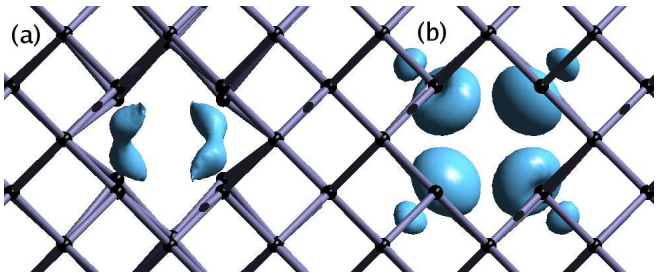}%
\caption{(Color online) The density of the localized electron state at
  a neutral Si vacancy when (a) there is no localized positron at the vacancy
  ($D_{2d}$ symmetry) and (b) a positron is localized at the vacancy
  (symmetric $T_{d}$).\label{deeplevel}}
\end{figure}

The strong effect of the trapped positron is seen also in the results 
for the vacancy point symmetry in semiconductors. Fig.~\ref{deeplevel}(a)
illustrates the case of the neutral $V_{\text{Si}}$. Without the 
trapped positron the Jahn-Teller effect lowers the point symmetry
to $D_{2d}$ and we see that the dangling bonds pointing toward the
center of the vacancy hybridize to two pairs of bonds between
the nearest-neighbor atoms. 
In Table \ref{energetics} the symmetry of the plain 
singly negative $V_{\text{Si}}$ is $D_{3d}$ corresponding to the 
split-vacancy configuration where one of the atoms neighboring the 
vacancy relaxes so that a divacancy with an atom in the center results.
These results for the Si vacancy are in good agreement with previous
LDA results.~\cite{Puska98,Wright06} 
With a positron localized at $V_{\text{Si}}$ the ideal lattice point 
symmetry $T_d$ is practically restored. As seen in Fig.~\ref{deeplevel}(b)
for the neutral $V_{\text{Si}}$ the strong positron repulsion increases 
the distances and weakens the bonds between the nearest-neighbor 
atoms of the vacancy 
and the atoms end up at the ideal-lattice point-symmetry positions 
within the numerical
accuracy. However, the deep localized electron state 
is not destroyed. This is the case also for the negative $V_{\text{Si}}$
and there will be ionization levels in the band gap also when the
vacancy traps a positron.

\subsubsection{Ionization levels}

\begin{figure}
\includegraphics[width=0.75\columnwidth]{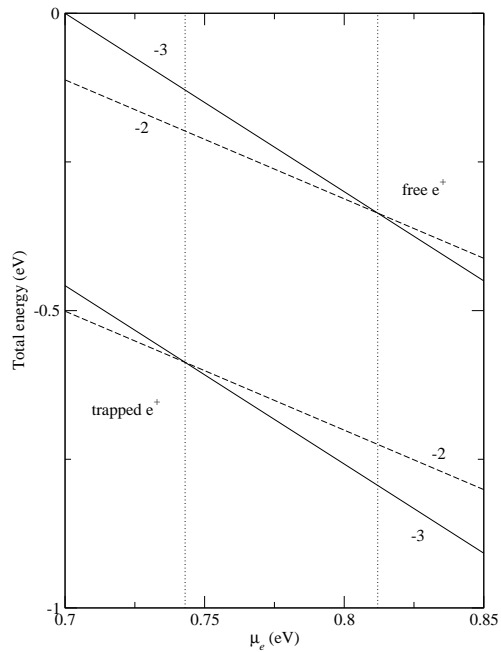}%
\caption{Total energy (zero level arbitrary) of a system comprised of
  a positron and the Ga vacancy in GaAs
  as function of the charge state of the vacancy and electron chemical
  potential $\mu_{e}$. The positron is either trapped at the
  vacancy or in the delocalized bulk state.\label{levels}}
\end{figure}

The positions of the thermodynamic ionization levels for vacancies in
semiconductors are also given in Table~\ref{energetics}.
Figure~\ref{levels} shows as an example the determination
of the level $\varepsilon$(--2/--3) 
for $V_{\text{Ga}}$ in GaAs without and with a trapped positron.
The position of the ionization level is given by the
point where the total energies cross. We see that
the introduction of the positron lowers the thermodynamical
ionization level. According to Table \ref{energetics}
the lowering of the ionization level due to positron
trapping is a general trend which reflects the lowering
of the positron energy eigenvalue at the vacancy due to the
excess negative charge. Slightly surprisingly,
the magnitude of the lowering is only of the order of 0.1 eV in spite
of the rather large ion relaxations 
due to the trapped positron. The small change is due to
the fact that the magnitudes of the ion relaxation in the
adjacent charge states before or after the positron trapping
are rather similar and because the excess electron density 
is rather delocalized and does not strongly lower the positron 
energy eigenvalue.

The behavior of the thermodynamic ionization levels
in relation to the positron trapping is an important result justifying 
positron experiments devoted for
determining ionization levels. Namely, we could think that the 
neutral and negatively charged states could become
thermodynamically unstable with respect to loosing 
a bound electron when a positron is trapped. However,
the lowering of the ionization levels indicates that this does not
occur. Actually, within our model the trapped positron
can then affect (lower) the measured ionization levels only in the
case of the level $\varepsilon(0/-)$ and maybe also 
in the case of the level $\varepsilon$(--/--2).
Namely, we expect that only the neutral or the singly negative
charge state (not counting the charge of the localized 
positron) can trap an electron due to the positron-induced
changes within the positron lifetime. The more negative charge
states 
effectively repel free electrons hindering their
trapping. The situation is similar to the positron trapping
at positively charged vacancies.~\cite{Puska90} On the other hand, the 
ionization level $\varepsilon(+/0)$ between 
the positive and the neutral charge state is determined
in the positron experiments by the fact that a positive defect 
does not trap a positron, and therefore the possible change of the
vacancy charge state from the neutral to the negative one does
not affect the determination of this level.

\subsubsection{Positron annihilation characteristics}

Positron lifetimes calculated for the different defects are
also given in Table~\ref{energetics}. The LDA for the
electron-positron enhancement effects calculated with the
Boro\'nski-Nieminen interpolation form underestimates the
positron lifetimes in comparison with experiment. The underestimation
is especially strong for materials containing $d$ electrons such as
transition metals and the III--V compound semiconductors such as
GaN. Also the lattice constants
calculated within the LDA for the electron exchange and 
correlation effects are too small compared with the 
measured ones decreasing the positron lifetimes. 
Therefore rather than the absolute lifetime values the 
differences or the ratios between the positron
defect and bulk lifetimes are the most important figures.
The ratios between the defect and bulk lifetimes are
according to Table \ref{energetics} about 1.1--1.25
and 1.5--1.8 for typical semiconductors and metals,
respectively. For $V_{\text{Ga}}$ in GaN and for $V_{\text{Mg}}$ the ratios
are 1.65 and 1.32, respectively. The ratios reflect the 
degree of the localization of the positron at the vacancies and, in
general, their trends are similar to the trends in the positron trapping
energies. It is interesting to note that the change
of the charge state to a more negative one slightly decreases 
the positron lifetime in typical semiconductors
whereas the lifetime in $V_{\text{Ga}}$ in GaN is insensitive to
the charge state. The deep-level electron wave functions in
$V_{\text{Ga}}$ in GaN are rather delocalized and adding more
electrons on the deep levels does not appreciably change the total
electron density and correspondingly the ionic relaxations.

\begin{figure}
\includegraphics[width=0.8\columnwidth]{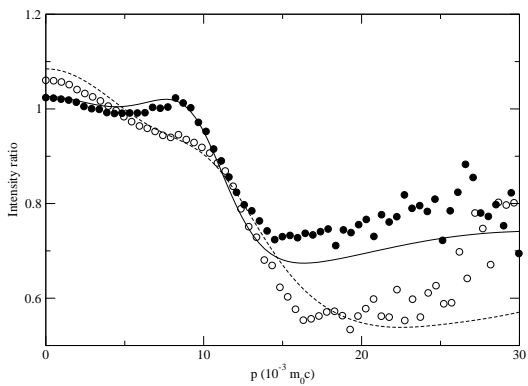}%
\caption{The experimental and calculated Doppler spectra (ratio to
  bulk) of triply negative Ga vacancies in GaN ($\circ$, dash line)
  and GaAs ($\bullet$, solid line). The experimental data are from
  Refs.~\onlinecite{Hautakangas06} and~\onlinecite{Laine96},
  respectively.\label{ratio}}
\end{figure}

Figure~\ref{ratio} shows ratio curves between the coincidence Doppler
broadening momentum distributions
of annihilating electron-positron pairs for vacancies and for the
corresponding bulk lattices. The measured and calculated curves
for triply-negative Ga vacancies in GaN and in GaAs are shown.
The data correspond to the [0001] and [001] directions. The
calculated distributions are convoluted with Gaussian functions
with the full width at the half maximum of 5.3 and 5.5 $\times 10^{-3}\ m_{0}c$
for GaN and GaAs, respectively, corresponding to the experimental
resolutions. The calculated curves quantitatively
reproduce the experimental trends. At low momenta the ratio for
GaN is higher than that for GaAs reflecting the larger reduction
of the electron density at the vacancy in GaN.
At high momenta the GaAs curve is above the GaN curve due
the contribution of As 3d electrons in GaAs. The agreement at high
momenta shows that our scheme is able to predict the overlap of
the positron and core electron densities or at least the relative
change in the positron-core electron overlap between the localized and
delocalized positron states.

To enable further studies of trends between different materials
Table~\ref{energetics} shows the relative $W$ parameters of vacancies,
$W_{\text{rel}}=W_{\text{defect}}/W_{\text{bulk}}$, that is an
experimental parameter reflecting the
decrease in core annihilation compared to the delocalized bulk state. Here
$W_{\text{defect}}$ is the $W$ parameter (integral over high-momentum
part of the Doppler spectrum) corresponding to the
localized defect state and $W_{\text{defect}}$ that of the delocalized
bulk state. The momentum window used is in all cases $15-30\times
10^{-3}\ m_{0}c$. The $W_{\text{rel}}$ parameter reflects the positron 
localization and the extent of the high-momentum core electrons
inside the vacancy. For example, it is interesting to note that the
$W_{\text{rel}}$ parameter is clearly smaller for Mg than for Al.

\section{Discussion and comparison with experiment}\label{discussion}


Our model gives for vacancies in typical metals and for cation vacancies
in compound semiconductors positron binding 
energies which are large enough that positron detrapping
at elevated temperatures even near the melting point is unlikely.
The predicted increases in the positron lifetimes in trapping
are in agreement with experimental values. For example,
for Al the calculated increase is 83 ps in agreement with the
experimental one of 85 ps (Ref.~\onlinecite{schaefer86})
and for for $V_{\text{Ga}}$ in GaN the calculated and
measured increases are 85 and 75 ps (Ref.~\onlinecite{Hautakangas06}),
respectively. The calculated and measured electron-positron momentum 
distributions at high momenta
also show a good correspondence (Fig.~\ref{ratio} gives an example) 
meaning that our model is able to reproduce the positron overlap with 
ion cores in a reasonable manner. 

The agreement between theory and experiment becomes less clear
when the predicted positron trapping energy decreases. In the
case of positron trapping at vacancies in Mg the calculated
positron lifetime increase is from 219 ps to 289 ps whereas the much
smaller increase from 225 ps to 255 ps has been measured between 
well a annealed sample and a sample with thermally generated
vacancies.~\cite{Hautojarvi82} 
The calculated large lifetime increase is also reflected in a 
rather small $W_{\text{rel}}$ parameter of 0.56 for 
$V_{\text{Mg}}$. However, the calculated positron trapping
energy of 0.34 eV is in agreement with the estimate of
$0.3-0.4$~eV by Hautoj\"arvi \textit{et al}.~\cite{Hautojarvi82}
One possible source of the theory-experiment disagreement
could be difficulties in extracting bulk and vacancy lifetimes
for Mg in measurements.



Our calculations predict a vanishingly small positron trapping energy at 
vacancies in Si, of the order of thermal energy at room
temperature. The trapped positron state does not exist when the
vacancy is relaxed without the influence of the localized positron,
which in principle prevents positron trapping at the vacancy at zero
temperature. Our result is, however, even qualitatively wrong since
the predicted trapping energy is negative.
However, the behavior of $V_{\text{Si}}$ during positron trapping
is unique reflecting the very flat energy landscape the ions
feel around the vacancy. The flatness is due to the different
competing possibilities for bonding and rebonding in a covalent
material. The flat energy landscape is also behind the scatter
of the DFT results for the structure and energetics of the Si vacancy.
Only recently, when calculations with very large supercells
(up to 1000 Si atoms) have become possible the results show
a satisfactory numerical convergence.~\cite{Wright06}
Figure~\ref{confcoord}(b) shows
that the energy landscape of the Si vacancy with a trapped positron 
is even flatter than that of the plain vacancy. This means,
as discussed above, that the entropy contribution should
be taken into account when describing the trapped state at
finite temperatures. However, one should bear in mind that
the errors arising, e.g., from the LDA's for the electron-positron 
correlation energy and for the electron-electron exchange and 
correlation may be of the order of tenths of an eV. For example,
the scatter in the calculated formation energies for $V_{\text{Si}}$
is of this order or even larger. 
Thus, our scheme may describe even qualitatively incorrectly
the actual positron trapping process in borderline cases such as
$V_{\text{Si}}$.

The flat energy landscape for $V_{\text{Si}}$ would introduce strong 
temperature-dependence to
positron trapping and detrapping processes. In our model
[Fig.~\ref{confcoord}(b)] the positron trapping would, in principle,
be possible at finite temperatures because part of the time the
vacancy volume is so large that a bound positron state exists. This
probability is, however, strongly temperature-dependent and in the
experiments the trapping rate as a function of temperature is seen to
be rather constant except for the case of negative defects for which
the trapping rate decreases with increasing temperature.~\cite{Puska90}

The concept of positron binding
energy seems to be a difficult one to define or at least it is
difficult to get a quantitative agreement between experiment and
theory even if the computational results were exact. First of all, one
of the 
assumptions behind the relation between trapping and detrapping
rates~\cite{Manninen81} typically used in the interpretation of
experimental data is that the excitations of the positron are
decoupled from the excitations of the
system. According to our calculation this clearly is not the
case. Secondly, the positron detrapping energy (the threshold energy
needed to detrap a localized positron) in the case of $V_{\text{Si}}$ strongly
depends on the detrapping mechanism (fast vertical transition vs slow
transition due to phonons) and on the ionic structure of the
defect at the instant of detrapping [see Fig.~\ref{confcoord}(b)]. In
the case of this kind of an energy landscape it is difficult to draw
conclusions about the actual positron trapping and detrapping
processes. According to our calculations the strain energy stored in the
relaxation of the vacancy is so large that all the released energy can
be stored in it (analogously to the trapping of a small
polaron). However, although the energy is conserved between the
initial and final states the
differing time scales in electronic (transition of the positron from
delocalized to localized state) and ionic processes (relaxation of the
vacancy during positron trapping) complicate the picture. Although our
results may not in all cases be even qualitatively correct they
clearly suggest that the models for positron trapping in
semiconductors~\cite{Puska90} need refinement.

Recent measurements of heavily As-doped Si indicate that
positron trapping energies at vacancy defects in highly As-doped Si may
be low and
thermal detrapping is possible.~\cite{Kuitunen07} Namely Kuitunen {\em et al.}\
found that positron detrapping happens from the Si vacancy decorated 
by three As atoms ($V_{\text{Si}}-{\text{As}}_3$) at temperatures above 500~K. 
For vacancies decorated with one or two As atoms they did not notice 
positron detrapping. Using the relation between the detrapping and trapping
rates derived by Manninen and Nieminen~\cite{Manninen81} Kuitunen {\em et al.}\
determined for $V_{\text{Si}}-{\text{As}}_3$ the trapping energy of 0.27~eV. 
Our scheme gives a clearly smaller positron trapping energy of
0.06--0.1 eV for $V_{\text{Si}}-{\text{As}}_3$. Similarly to the case
of $V_{\text{Si}}$ we do not find for
$V_{\text{Si}}-{\text{As}}$ and $V_{\text{Si}}-{\text{As}}_2$ an energetically
favored trapped positron state but just a metastable configuration in
disagreement with the experimental
trend. The trend in our trapping energies is exactly the opposite; the
trapping energy increases with the increasing number of As atoms
($n=0,...,3$). This is because the large As ions around the vacancy
do not relax inwards as strongly as the neighboring Si ions so that
the energy stored in the ionic lattice in the positron trapping process
is smaller for the As decorated vacancy than for the clean vacancy.
A well-known shallow positron trap in Si with open volume is the
complex formed by a vacancy and an interstitial oxygen. For it, the reported
positron binding energies are of the order of 40--50~meV
(Refs.~\onlinecite{Polity98,Kauppinen99}).
The open volume of the $V-\text{O}$ complex is so small (see
Ref.~\onlinecite{Pesola99}) that we do not expect it to trap positrons
in our calculations.

The fact that a bound positron state can be found at $V_{\text{Ge}}$
or at $V_{\text{As}}$ in GaAs when freezing the ions at ideal positions
means that the bound positron states are very close to appear, and
an improvement in the theoretical description could lead to bound
positron states also for optimized ion positions. In experimental
works~\cite{Polity99,Wurschum89,Moser85} positron lifetime
components between  279 and 292~ps are assigned 
to $V_{\text{Ge}}$. The measured lifetimes can be contrasted to the
measured bulk lifetime of 228~ps (Ref.~\onlinecite{Polity99}). These vacancy
lifetimes are already quite close to the theoretical estimate of 316
ps for an ideal divacancy in Ge (the corresponding bulk lifetime is
229~ps).~\cite{Puska89} For an ideal neutral 
monovacancy we get the lifetime of 246~ps which is only 33 ps longer
than our bulk lifetime. These comparisons suggest that the
experimentally observed lifetime components are too long to be explained by
annihilation at ideal monovacancy-size defects.
Measurements~\cite{Saarinen91a} for $n$-type GaAs show the lifetimes
of 257 ps and 295 ps (the experimental bulk lifetime is 231 ps).
These were assigned to negative and neutral As vacancies possibly
associated with impurity atoms on the basis of measured ionization
levels and corresponding old computational
results~\cite{Puska89b,Xu90} which suggested that the levels of
$V_{\text{As}}$ are near the conduction band. The measurements
gave ionization levels of $\varepsilon(+/0)$ =
$E_c$ -- 0.140 eV and $\varepsilon(0/-)$ = $E_c$ -- 0.030 eV, where
$E_{c}$ is the
conductance band minimum. Our first-principles
results, which are in accordance with those in 
Ref.~\onlinecite{El-Mellouhi05}, indicate that these ionization levels
of clean $V_{\text{As}}$ in GaAs are close to the top of the valence band.
This supports also the conclusions that the defects observed in positron
measurements are not clean vacancies.
 
Finally, our calculations for  $V_{\text{N}}$ in GaN suggest that
bound positron states at small anion vacancies in compound 
semiconductors are not possible. This should be contrasted with
a positron annihilation study in which the detection of N vacancies is
reported.~\cite{Hautakangas03} In the interpretation, however, the
short lifetime component is associated with $V_{\text{N}}$-impurity
complexes and not with isolated $V_{\text{N}}$.

\section{Conclusions}\label{conclusions}

We have studied, using first-principles calculations, the role of
lattice relaxations around vacancies in the positron trapping process
in various metals and semiconductors. In metals the trapping
energies are typically of the order of 1\ldots 2 eV. The lattice relaxes
due to the trapped positron and the positron annihilation parameters
change but especially the qualitative effects are small. The most
important difference
between typical metals and semiconductors is the magnitude of the strain
energy stored in the lattice compared to the lowering of the positron
energy eigenvalue. For semiconductors these two can be of the same
order of magnitude because of the smaller reduction of nucleus repulsion
in the trapping process. This leads to rather small values for the
trapping energy.

In the case of covalently bonded elemental semiconductors such as Si
and Ge the energy landscape of the positron-vacancy system is
extremely flat which suggests that entropic considerations have to be
taken into account when discussing the ionic structure of the
vacancy with a localized positron at finite temperatures. For Si the
calculations give, in disagreement with experiments, vanishingly small
or rather sightly negative trapping energies 
suggesting that trapping of thermalized positrons would not even be
energetically favorable.

Also the vacancy in Ge and anion vacancies in
compound semiconductors are challenging tests for
theoretical methods since in the calculations there is either no bound
positron state at the vacancy or the trapping is energetically
unfavorable and there is no local energy minimum configuration at
which the positron is trapped at the vacancy.
Furthermore, for the vacancy in Si the configuration with the trapped
positron is just a metastable state while in the ground state the
positron is in the delocalized bulk state.
An interesting finding is that a localized positron cancels
all the Jahn-Teller distortions we observed when having no positron at
the vacancy.

In general, the higher the predicted positron trapping energy is the
better is the agreement between our results and experiments. However,
also in the case of defects in Si we get a surprisingly good agreement
in calculated lifetimes and Doppler broadening spectra 
for the metastable state.

We have also studied the effect of the localized positron on the
electronic structure of the vacancies in semiconductors by evaluating
thermodynamical ionization levels of vacancies. The positron-induced
changes are usually only of the order of 0.1~eV. In general, the levels
move closer to the valence band maximum. A positron-induced change in
the defect charge state in this model is likely only in the case of a
neutral defect since positive ones do not trap positrons and negative
ones electrons within the lifetime of the trapped positron.


\begin{acknowledgments}
We are grateful for discussions with Academy Prof.\ R.\ M.\ Nieminen and
Dr.\ F.\ Tuomisto. We also acknowledge the
generous computer resources from the Center of Scientific Computing,
Espoo, Finland. I.M.\ acknowledges the financial support by the
Finnish Academy of Science and Letters, Vilho,
Yrj\"o, and Kalle V\"ais\"al\"a Foundation.
\end{acknowledgments}


\begin{thebibliography}{54}
\expandafter\ifx\csname natexlab\endcsname\relax\def\natexlab#1{#1}\fi
\expandafter\ifx\csname bibnamefont\endcsname\relax
  \def\bibnamefont#1{#1}\fi
\expandafter\ifx\csname bibfnamefont\endcsname\relax
  \def\bibfnamefont#1{#1}\fi
\expandafter\ifx\csname citenamefont\endcsname\relax
  \def\citenamefont#1{#1}\fi
\expandafter\ifx\csname url\endcsname\relax
  \def\url#1{\texttt{#1}}\fi
\expandafter\ifx\csname urlprefix\endcsname\relax\def\urlprefix{URL }\fi
\providecommand{\bibinfo}[2]{#2}
\providecommand{\eprint}[2][]{\url{#2}}

\bibitem[{\citenamefont{Krause-Rehberg and Leipner}(1999)}]{Krause-Rehberg99}
\bibinfo{author}{\bibfnamefont{R.}~\bibnamefont{Krause-Rehberg}}
  \bibnamefont{and} \bibinfo{author}{\bibfnamefont{H.}~\bibnamefont{Leipner}},
  \emph{\bibinfo{title}{Positron Annihilation in Semiconductors}}
  (\bibinfo{publisher}{Springer-Verlag}, \bibinfo{address}{Berlin},
  \bibinfo{year}{1999}).

\bibitem[{\citenamefont{Puska and Nieminen}(1994)}]{Puska94}
\bibinfo{author}{\bibfnamefont{M.~J.} \bibnamefont{Puska}} \bibnamefont{and}
  \bibinfo{author}{\bibfnamefont{R.~M.} \bibnamefont{Nieminen}},
  \bibinfo{journal}{Rev. Mod. Phys.} \textbf{\bibinfo{volume}{66}},
  \bibinfo{pages}{841} (\bibinfo{year}{1994}).

\bibitem[{\citenamefont{Hodges}(1970)}]{Hodges70}
\bibinfo{author}{\bibfnamefont{C.~H.} \bibnamefont{Hodges}},
  \bibinfo{journal}{Phys. Rev. Lett.} \textbf{\bibinfo{volume}{25}},
  \bibinfo{pages}{284} (\bibinfo{year}{1970}).

\bibitem[{\citenamefont{Nieminen and Laakkonen}(1979)}]{Nieminen79}
\bibinfo{author}{\bibfnamefont{R.~M.} \bibnamefont{Nieminen}} \bibnamefont{and}
  \bibinfo{author}{\bibfnamefont{J.}~\bibnamefont{Laakkonen}},
  \bibinfo{journal}{Appl. Phys} \textbf{\bibinfo{volume}{20}},
  \bibinfo{pages}{181} (\bibinfo{year}{1979}).

\bibitem[{\citenamefont{Puska et~al.}(1990)\citenamefont{Puska, Corbel, and
  Nieminen}}]{Puska90}
\bibinfo{author}{\bibfnamefont{M.~J.} \bibnamefont{Puska}},
  \bibinfo{author}{\bibfnamefont{C.}~\bibnamefont{Corbel}}, \bibnamefont{and}
  \bibinfo{author}{\bibfnamefont{R.~M.} \bibnamefont{Nieminen}},
  \bibinfo{journal}{Phys. Rev. B} \textbf{\bibinfo{volume}{41}},
  \bibinfo{pages}{9980} (\bibinfo{year}{1990}).

\bibitem[{\citenamefont{Corbel et~al.}(1988)\citenamefont{Corbel, Stucky,
  Hautoj\"arvi, Saarinen, and Moser}}]{Corbel88}
\bibinfo{author}{\bibfnamefont{C.}~\bibnamefont{Corbel}},
  \bibinfo{author}{\bibfnamefont{M.}~\bibnamefont{Stucky}},
  \bibinfo{author}{\bibfnamefont{P.}~\bibnamefont{Hautoj\"arvi}},
  \bibinfo{author}{\bibfnamefont{K.}~\bibnamefont{Saarinen}}, \bibnamefont{and}
  \bibinfo{author}{\bibfnamefont{P.}~\bibnamefont{Moser}},
  \bibinfo{journal}{Phys. Rev. B} \textbf{\bibinfo{volume}{38}},
  \bibinfo{pages}{8192} (\bibinfo{year}{1988}).

\bibitem[{\citenamefont{Saarinen et~al.}(1991)\citenamefont{Saarinen,
  Hautoj\"arvi, Lanki, and Corbel}}]{Saarinen91a}
\bibinfo{author}{\bibfnamefont{K.}~\bibnamefont{Saarinen}},
  \bibinfo{author}{\bibfnamefont{P.}~\bibnamefont{Hautoj\"arvi}},
  \bibinfo{author}{\bibfnamefont{P.}~\bibnamefont{Lanki}}, \bibnamefont{and}
  \bibinfo{author}{\bibfnamefont{C.}~\bibnamefont{Corbel}},
  \bibinfo{journal}{Phys. Rev. B} \textbf{\bibinfo{volume}{44}},
  \bibinfo{pages}{10585} (\bibinfo{year}{1991}).

\bibitem[{\citenamefont{Saarinen et~al.}(1993)\citenamefont{Saarinen, Kuisma,
  Hautoj\"arvi, Corbel, and LeBerre}}]{Saarinen91b}
\bibinfo{author}{\bibfnamefont{K.}~\bibnamefont{Saarinen}},
  \bibinfo{author}{\bibfnamefont{S.}~\bibnamefont{Kuisma}},
  \bibinfo{author}{\bibfnamefont{P.}~\bibnamefont{Hautoj\"arvi}},
  \bibinfo{author}{\bibfnamefont{C.}~\bibnamefont{Corbel}}, \bibnamefont{and}
  \bibinfo{author}{\bibfnamefont{C.}~\bibnamefont{LeBerre}},
  \bibinfo{journal}{Phys. Rev. Lett.} \textbf{\bibinfo{volume}{70}},
  \bibinfo{pages}{2794} (\bibinfo{year}{1993}).

\bibitem[{\citenamefont{LeBerre et~al.}(1995)\citenamefont{LeBerre, Corbel,
  Mih, Brosel, T\"uzemen, Kuisma, Saarinen, Hautoj\"arvi, and
  Fornari}}]{LeBerre95}
\bibinfo{author}{\bibfnamefont{C.}~\bibnamefont{LeBerre}},
  \bibinfo{author}{\bibfnamefont{C.}~\bibnamefont{Corbel}},
  \bibinfo{author}{\bibfnamefont{R.}~\bibnamefont{Mih}},
  \bibinfo{author}{\bibfnamefont{M.~R.} \bibnamefont{Brosel}},
  \bibinfo{author}{\bibfnamefont{S.}~\bibnamefont{T\"uzemen}},
  \bibinfo{author}{\bibfnamefont{S.}~\bibnamefont{Kuisma}},
  \bibinfo{author}{\bibfnamefont{K.}~\bibnamefont{Saarinen}},
  \bibinfo{author}{\bibfnamefont{P.}~\bibnamefont{Hautoj\"arvi}},
  \bibnamefont{and} \bibinfo{author}{\bibfnamefont{R.}~\bibnamefont{Fornari}},
  \bibinfo{journal}{Appl. Phys. Lett.} \textbf{\bibinfo{volume}{66}},
  \bibinfo{pages}{2534} (\bibinfo{year}{1995}).

\bibitem[{\citenamefont{Kuisma et~al.}(1997)\citenamefont{Kuisma, Saarinen,
  Hautoj\"arvi, and Corbel}}]{Kuisma97}
\bibinfo{author}{\bibfnamefont{S.}~\bibnamefont{Kuisma}},
  \bibinfo{author}{\bibfnamefont{K.}~\bibnamefont{Saarinen}},
  \bibinfo{author}{\bibfnamefont{P.}~\bibnamefont{Hautoj\"arvi}},
  \bibnamefont{and} \bibinfo{author}{\bibfnamefont{C.}~\bibnamefont{Corbel}},
  \bibinfo{journal}{Phys. Rev. B} \textbf{\bibinfo{volume}{55}},
  \bibinfo{pages}{9609} (\bibinfo{year}{1997}).

\bibitem[{\citenamefont{Kauppinen et~al.}(1998)\citenamefont{Kauppinen, Corbel,
  Nissil\"a, Saarinen, and Hautoj\"arvi}}]{Kauppinen99}
\bibinfo{author}{\bibfnamefont{H.}~\bibnamefont{Kauppinen}},
  \bibinfo{author}{\bibfnamefont{C.}~\bibnamefont{Corbel}},
  \bibinfo{author}{\bibfnamefont{J.}~\bibnamefont{Nissil\"a}},
  \bibinfo{author}{\bibfnamefont{K.}~\bibnamefont{Saarinen}}, \bibnamefont{and}
  \bibinfo{author}{\bibfnamefont{P.}~\bibnamefont{Hautoj\"arvi}},
  \bibinfo{journal}{Phys. Rev. B} \textbf{\bibinfo{volume}{57}},
  \bibinfo{pages}{12911} (\bibinfo{year}{1998}).

\bibitem[{\citenamefont{Arpiainen et~al.}(2002)\citenamefont{Arpiainen,
  Saarinen, Hautoj\"arvi, Henry, Barthe, and Corbel}}]{Arpiainen02}
\bibinfo{author}{\bibfnamefont{S.}~\bibnamefont{Arpiainen}},
  \bibinfo{author}{\bibfnamefont{K.}~\bibnamefont{Saarinen}},
  \bibinfo{author}{\bibfnamefont{P.}~\bibnamefont{Hautoj\"arvi}},
  \bibinfo{author}{\bibfnamefont{L.}~\bibnamefont{Henry}},
  \bibinfo{author}{\bibfnamefont{M.-F.} \bibnamefont{Barthe}},
  \bibnamefont{and} \bibinfo{author}{\bibfnamefont{C.}~\bibnamefont{Corbel}},
  \bibinfo{journal}{Phys. Rev. B} \textbf{\bibinfo{volume}{66}},
  \bibinfo{pages}{075206} (\bibinfo{year}{2002}).

\bibitem[{\citenamefont{Tuomisto et~al.}(2005)\citenamefont{Tuomisto, Saarinen,
  Look, and Farlow}}]{Tuomisto05}
\bibinfo{author}{\bibfnamefont{F.}~\bibnamefont{Tuomisto}},
  \bibinfo{author}{\bibfnamefont{K.}~\bibnamefont{Saarinen}},
  \bibinfo{author}{\bibfnamefont{D.~C.} \bibnamefont{Look}}, \bibnamefont{and}
  \bibinfo{author}{\bibfnamefont{G.~C.} \bibnamefont{Farlow}},
  \bibinfo{journal}{Phys. Rev. B} \textbf{\bibinfo{volume}{72}},
  \bibinfo{pages}{085206} (\bibinfo{year}{2005}).

\bibitem[{\citenamefont{M\"akinen et~al.}(1992)\citenamefont{M\"akinen,
  Hautoj\"arvi, and Corbel}}]{Makinen92}
\bibinfo{author}{\bibfnamefont{J.}~\bibnamefont{M\"akinen}},
  \bibinfo{author}{\bibfnamefont{P.}~\bibnamefont{Hautoj\"arvi}},
  \bibnamefont{and} \bibinfo{author}{\bibfnamefont{C.}~\bibnamefont{Corbel}},
  \bibinfo{journal}{J. Phys.: Condens. Matter} \textbf{\bibinfo{volume}{4}},
  \bibinfo{pages}{5137} (\bibinfo{year}{1992}).

\bibitem[{\citenamefont{Puska and Nieminen}(1983)}]{Puska83}
\bibinfo{author}{\bibfnamefont{M.~J.} \bibnamefont{Puska}} \bibnamefont{and}
  \bibinfo{author}{\bibfnamefont{R.~M.} \bibnamefont{Nieminen}},
  \bibinfo{journal}{J. Phys. F} \textbf{\bibinfo{volume}{13}},
  \bibinfo{pages}{333} (\bibinfo{year}{1983}).

\bibitem[{\citenamefont{Puska et~al.}(1989)\citenamefont{Puska, M\"akinen,
  Manninen, and Nieminen}}]{Puska89}
\bibinfo{author}{\bibfnamefont{M.~J.} \bibnamefont{Puska}},
  \bibinfo{author}{\bibfnamefont{S.}~\bibnamefont{M\"akinen}},
  \bibinfo{author}{\bibfnamefont{M.}~\bibnamefont{Manninen}}, \bibnamefont{and}
  \bibinfo{author}{\bibfnamefont{R.~M.} \bibnamefont{Nieminen}},
  \bibinfo{journal}{Phys. Rev. B} \textbf{\bibinfo{volume}{39}},
  \bibinfo{pages}{7666} (\bibinfo{year}{1989}).

\bibitem[{\citenamefont{Barbiellini et~al.}(1996)\citenamefont{Barbiellini,
  Puska, Korhonen, Harju, Torsti, and Nieminen}}]{Barbiellini96}
\bibinfo{author}{\bibfnamefont{B.}~\bibnamefont{Barbiellini}},
  \bibinfo{author}{\bibfnamefont{M.~J.} \bibnamefont{Puska}},
  \bibinfo{author}{\bibfnamefont{T.}~\bibnamefont{Korhonen}},
  \bibinfo{author}{\bibfnamefont{A.}~\bibnamefont{Harju}},
  \bibinfo{author}{\bibfnamefont{T.}~\bibnamefont{Torsti}}, \bibnamefont{and}
  \bibinfo{author}{\bibfnamefont{R.~M.} \bibnamefont{Nieminen}},
  \bibinfo{journal}{Phys Rev. B} \textbf{\bibinfo{volume}{53}},
  \bibinfo{pages}{16201} (\bibinfo{year}{1996}).

\bibitem[{\citenamefont{Makkonen et~al.}(2006)\citenamefont{Makkonen, Hakala,
  and Puska}}]{Makkonen06}
\bibinfo{author}{\bibfnamefont{I.}~\bibnamefont{Makkonen}},
  \bibinfo{author}{\bibfnamefont{M.}~\bibnamefont{Hakala}}, \bibnamefont{and}
  \bibinfo{author}{\bibfnamefont{M.~J.} \bibnamefont{Puska}},
  \bibinfo{journal}{Phys. Rev. B} \textbf{\bibinfo{volume}{73}},
  \bibinfo{pages}{035103} (\bibinfo{year}{2006}).

\bibitem[{\citenamefont{Hohenberg and Kohn}(1964)}]{Hohenberg64}
\bibinfo{author}{\bibfnamefont{P.}~\bibnamefont{Hohenberg}} \bibnamefont{and}
  \bibinfo{author}{\bibfnamefont{W.}~\bibnamefont{Kohn}},
  \bibinfo{journal}{Phys. Rev.} \textbf{\bibinfo{volume}{136}},
  \bibinfo{pages}{B864} (\bibinfo{year}{1964}).

\bibitem[{\citenamefont{Kohn and Sham}(1965)}]{Kohn65}
\bibinfo{author}{\bibfnamefont{W.}~\bibnamefont{Kohn}} \bibnamefont{and}
  \bibinfo{author}{\bibfnamefont{J.}~\bibnamefont{Sham}},
  \bibinfo{journal}{Phys. Rev.} \textbf{\bibinfo{volume}{140}},
  \bibinfo{pages}{A1133} (\bibinfo{year}{1965}).

\bibitem[{\citenamefont{Boro\'nski and Nieminen}(1986)}]{Boronski86}
\bibinfo{author}{\bibfnamefont{E.}~\bibnamefont{Boro\'nski}} \bibnamefont{and}
  \bibinfo{author}{\bibfnamefont{R.~M.} \bibnamefont{Nieminen}},
  \bibinfo{journal}{Phys. Rev. B} \textbf{\bibinfo{volume}{34}},
  \bibinfo{pages}{3820} (\bibinfo{year}{1986}).

\bibitem[{\citenamefont{Gilgien et~al.}(1994)\citenamefont{Gilgien, Galli,
  Gygi, and Car}}]{Gilgien94}
\bibinfo{author}{\bibfnamefont{L.}~\bibnamefont{Gilgien}},
  \bibinfo{author}{\bibfnamefont{G.}~\bibnamefont{Galli}},
  \bibinfo{author}{\bibfnamefont{F.}~\bibnamefont{Gygi}}, \bibnamefont{and}
  \bibinfo{author}{\bibfnamefont{R.}~\bibnamefont{Car}},
  \bibinfo{journal}{Phys.\ Rev.\ Lett.} \textbf{\bibinfo{volume}{72}},
  \bibinfo{pages}{3214} (\bibinfo{year}{1994}).

\bibitem[{\citenamefont{Puska et~al.}(1995)\citenamefont{Puska, Seitsonen, and
  Nieminen}}]{Puska95}
\bibinfo{author}{\bibfnamefont{M.~J.} \bibnamefont{Puska}},
  \bibinfo{author}{\bibfnamefont{A.~P.} \bibnamefont{Seitsonen}},
  \bibnamefont{and} \bibinfo{author}{\bibfnamefont{R.~M.}
  \bibnamefont{Nieminen}}, \bibinfo{journal}{Phys.\ Rev.\ B}
  \textbf{\bibinfo{volume}{52}}, \bibinfo{pages}{10947} (\bibinfo{year}{1995}).

\bibitem[{\citenamefont{Saito and Oshiyama}(1996)}]{Saito96}
\bibinfo{author}{\bibfnamefont{M.}~\bibnamefont{Saito}} \bibnamefont{and}
  \bibinfo{author}{\bibfnamefont{A.}~\bibnamefont{Oshiyama}},
  \bibinfo{journal}{Phys.\ Rev.\ B} \textbf{\bibinfo{volume}{53}},
  \bibinfo{pages}{7810} (\bibinfo{year}{1996}).

\bibitem[{\citenamefont{Tang et~al.}(1997)\citenamefont{Tang, Hasegawa, Chiba,
  Saito, Kawasuso, Li, Fu, Akahane, Kawazoe, and Yamaguchi}}]{Tang97}
\bibinfo{author}{\bibfnamefont{Z.}~\bibnamefont{Tang}},
  \bibinfo{author}{\bibfnamefont{M.}~\bibnamefont{Hasegawa}},
  \bibinfo{author}{\bibfnamefont{T.}~\bibnamefont{Chiba}},
  \bibinfo{author}{\bibfnamefont{M.}~\bibnamefont{Saito}},
  \bibinfo{author}{\bibfnamefont{A.}~\bibnamefont{Kawasuso}},
  \bibinfo{author}{\bibfnamefont{Z.~Q.} \bibnamefont{Li}},
  \bibinfo{author}{\bibfnamefont{R.~T.} \bibnamefont{Fu}},
  \bibinfo{author}{\bibfnamefont{T.}~\bibnamefont{Akahane}},
  \bibinfo{author}{\bibfnamefont{Y.}~\bibnamefont{Kawazoe}}, \bibnamefont{and}
  \bibinfo{author}{\bibfnamefont{S.}~\bibnamefont{Yamaguchi}},
  \bibinfo{journal}{Phys. Rev. Lett.} \textbf{\bibinfo{volume}{78}},
  \bibinfo{pages}{2236} (\bibinfo{year}{1997}).

\bibitem[{\citenamefont{Makhov and Lewis}(2005)}]{Makhov05}
\bibinfo{author}{\bibfnamefont{D.~V.} \bibnamefont{Makhov}} \bibnamefont{and}
  \bibinfo{author}{\bibfnamefont{L.~J.} \bibnamefont{Lewis}},
  \bibinfo{journal}{Phys.\ Rev.\ B} \textbf{\bibinfo{volume}{71}},
  \bibinfo{pages}{205215} (\bibinfo{year}{2005}).

\bibitem[{\citenamefont{Alatalo et~al.}(1996)\citenamefont{Alatalo,
  Barbiellini, Hakala, Kauppinen, Korhonen, Puska, Saarinen, Hautoj\"arvi, and
  Nieminen}}]{Alatalo96}
\bibinfo{author}{\bibfnamefont{M.}~\bibnamefont{Alatalo}},
  \bibinfo{author}{\bibfnamefont{B.}~\bibnamefont{Barbiellini}},
  \bibinfo{author}{\bibfnamefont{M.}~\bibnamefont{Hakala}},
  \bibinfo{author}{\bibfnamefont{H.}~\bibnamefont{Kauppinen}},
  \bibinfo{author}{\bibfnamefont{T.}~\bibnamefont{Korhonen}},
  \bibinfo{author}{\bibfnamefont{M.~J.} \bibnamefont{Puska}},
  \bibinfo{author}{\bibfnamefont{K.}~\bibnamefont{Saarinen}},
  \bibinfo{author}{\bibfnamefont{P.}~\bibnamefont{Hautoj\"arvi}},
  \bibnamefont{and} \bibinfo{author}{\bibfnamefont{R.~M.}
  \bibnamefont{Nieminen}}, \bibinfo{journal}{Phys. Rev. B}
  \textbf{\bibinfo{volume}{54}}, \bibinfo{pages}{2397} (\bibinfo{year}{1996}).

\bibitem[{\citenamefont{Barbiellini et~al.}(1997)\citenamefont{Barbiellini,
  Hakala, Puska, Nieminen, and Manuel}}]{Barbiellini97}
\bibinfo{author}{\bibfnamefont{B.}~\bibnamefont{Barbiellini}},
  \bibinfo{author}{\bibfnamefont{M.}~\bibnamefont{Hakala}},
  \bibinfo{author}{\bibfnamefont{M.~J.} \bibnamefont{Puska}},
  \bibinfo{author}{\bibfnamefont{R.~M.} \bibnamefont{Nieminen}},
  \bibnamefont{and} \bibinfo{author}{\bibfnamefont{A.~A.}
  \bibnamefont{Manuel}}, \bibinfo{journal}{Phys. Rev. B}
  \textbf{\bibinfo{volume}{56}}, \bibinfo{pages}{7136} (\bibinfo{year}{1997}).

\bibitem[{\citenamefont{Makov and Payne}(1995)}]{Makov95}
\bibinfo{author}{\bibfnamefont{G.}~\bibnamefont{Makov}} \bibnamefont{and}
  \bibinfo{author}{\bibfnamefont{M.~C.} \bibnamefont{Payne}},
  \bibinfo{journal}{Phys. Rev. B} \textbf{\bibinfo{volume}{51}},
  \bibinfo{pages}{4014} (\bibinfo{year}{1995}).

\bibitem[{\citenamefont{Perdew and Zunger}(1981)}]{Perdew81}
\bibinfo{author}{\bibfnamefont{J.~P.} \bibnamefont{Perdew}} \bibnamefont{and}
  \bibinfo{author}{\bibfnamefont{A.}~\bibnamefont{Zunger}},
  \bibinfo{journal}{Phys. Rev. B} \textbf{\bibinfo{volume}{23}},
  \bibinfo{pages}{5048} (\bibinfo{year}{1981}).

\bibitem[{\citenamefont{Bl\"ochl}(1994)}]{PAW}
\bibinfo{author}{\bibfnamefont{P.~E.} \bibnamefont{Bl\"ochl}},
  \bibinfo{journal}{Phys. Rev. B} \textbf{\bibinfo{volume}{50}},
  \bibinfo{pages}{17953} (\bibinfo{year}{1994}).

\bibitem[{\citenamefont{Kresse and
  Furthm\"uller}(1996{\natexlab{a}})}]{Kresse96a}
\bibinfo{author}{\bibfnamefont{G.}~\bibnamefont{Kresse}} \bibnamefont{and}
  \bibinfo{author}{\bibfnamefont{J.}~\bibnamefont{Furthm\"uller}},
  \bibinfo{journal}{Comput. Mat. Sci.} \textbf{\bibinfo{volume}{6}},
  \bibinfo{pages}{15} (\bibinfo{year}{1996}{\natexlab{a}}).

\bibitem[{\citenamefont{Kresse and
  Furthm\"uller}(1996{\natexlab{b}})}]{Kresse96b}
\bibinfo{author}{\bibfnamefont{G.}~\bibnamefont{Kresse}} \bibnamefont{and}
  \bibinfo{author}{\bibfnamefont{J.}~\bibnamefont{Furthm\"uller}},
  \bibinfo{journal}{Phys. Rev. B} \textbf{\bibinfo{volume}{54}},
  \bibinfo{pages}{11169} (\bibinfo{year}{1996}{\natexlab{b}}).

\bibitem[{\citenamefont{Kresse and Joubert}(1999)}]{PAWKresse}
\bibinfo{author}{\bibfnamefont{G.}~\bibnamefont{Kresse}} \bibnamefont{and}
  \bibinfo{author}{\bibfnamefont{D.}~\bibnamefont{Joubert}},
  \bibinfo{journal}{Phys. Rev. B} \textbf{\bibinfo{volume}{59}},
  \bibinfo{pages}{1758} (\bibinfo{year}{1999}).

\bibitem[{\citenamefont{Monkhorst and Pack}(1976)}]{Monkhorst76}
\bibinfo{author}{\bibfnamefont{H.~J.} \bibnamefont{Monkhorst}}
  \bibnamefont{and} \bibinfo{author}{\bibfnamefont{J.~D.} \bibnamefont{Pack}},
  \bibinfo{journal}{Phys. Rev. B} \textbf{\bibinfo{volume}{13}},
  \bibinfo{pages}{5188} (\bibinfo{year}{1976}).

\bibitem[{\citenamefont{Coutinho et~al.}(2006)\citenamefont{Coutinho, \"Oberg,
  Torres, Barroso, Jones, and Briddon}}]{Coutinho06}
\bibinfo{author}{\bibfnamefont{J.}~\bibnamefont{Coutinho}},
  \bibinfo{author}{\bibfnamefont{S.}~\bibnamefont{\"Oberg}},
  \bibinfo{author}{\bibfnamefont{V.~J.~B.} \bibnamefont{Torres}},
  \bibinfo{author}{\bibfnamefont{M.}~\bibnamefont{Barroso}},
  \bibinfo{author}{\bibfnamefont{R.}~\bibnamefont{Jones}}, \bibnamefont{and}
  \bibinfo{author}{\bibfnamefont{P.~R.} \bibnamefont{Briddon}},
  \bibinfo{journal}{Phys. Rev. B} \textbf{\bibinfo{volume}{73}},
  \bibinfo{pages}{235213} (\bibinfo{year}{2006}).

\bibitem[{\citenamefont{Mott and Stoneham}(1977)}]{Mott77}
\bibinfo{author}{\bibfnamefont{N.~F.} \bibnamefont{Mott}} \bibnamefont{and}
  \bibinfo{author}{\bibfnamefont{A.~M.} \bibnamefont{Stoneham}},
  \bibinfo{journal}{J. Phys. C} \textbf{\bibinfo{volume}{10}},
  \bibinfo{pages}{3391} (\bibinfo{year}{1977}).

\bibitem[{\citenamefont{Puska et~al.}(1998)\citenamefont{Puska, P\"oykk\"o,
  Pesola, and Nieminen}}]{Puska98}
\bibinfo{author}{\bibfnamefont{M.~J.} \bibnamefont{Puska}},
  \bibinfo{author}{\bibfnamefont{S.}~\bibnamefont{P\"oykk\"o}},
  \bibinfo{author}{\bibfnamefont{M.}~\bibnamefont{Pesola}}, \bibnamefont{and}
  \bibinfo{author}{\bibfnamefont{R.~M.} \bibnamefont{Nieminen}},
  \bibinfo{journal}{Phys.\ Rev.\ B} \textbf{\bibinfo{volume}{58}},
  \bibinfo{pages}{1318} (\bibinfo{year}{1998}).

\bibitem[{\citenamefont{Wright}(2006)}]{Wright06}
\bibinfo{author}{\bibfnamefont{A.~F.} \bibnamefont{Wright}},
  \bibinfo{journal}{Phys. Rev. B} \textbf{\bibinfo{volume}{74}},
  \bibinfo{pages}{165116} (\bibinfo{year}{2006}).

\bibitem[{\citenamefont{Hautakangas et~al.}(2006)\citenamefont{Hautakangas,
  Makkonen, Ranki, Puska, Saarinen, Xu, and Look}}]{Hautakangas06}
\bibinfo{author}{\bibfnamefont{S.}~\bibnamefont{Hautakangas}},
  \bibinfo{author}{\bibfnamefont{I.}~\bibnamefont{Makkonen}},
  \bibinfo{author}{\bibfnamefont{V.}~\bibnamefont{Ranki}},
  \bibinfo{author}{\bibfnamefont{M.~J.} \bibnamefont{Puska}},
  \bibinfo{author}{\bibfnamefont{K.}~\bibnamefont{Saarinen}},
  \bibinfo{author}{\bibfnamefont{X.}~\bibnamefont{Xu}}, \bibnamefont{and}
  \bibinfo{author}{\bibfnamefont{D.~C.} \bibnamefont{Look}},
  \bibinfo{journal}{Phys. Rev. B} \textbf{\bibinfo{volume}{73}},
  \bibinfo{pages}{193301} (\bibinfo{year}{2006}).

\bibitem[{\citenamefont{Laine et~al.}(1996)\citenamefont{Laine, Saarinen,
  M\"akinen, Hautoj\"arvi, Corbel, Pfeiffer, and Citrin}}]{Laine96}
\bibinfo{author}{\bibfnamefont{T.}~\bibnamefont{Laine}},
  \bibinfo{author}{\bibfnamefont{K.}~\bibnamefont{Saarinen}},
  \bibinfo{author}{\bibfnamefont{J.}~\bibnamefont{M\"akinen}},
  \bibinfo{author}{\bibfnamefont{P.}~\bibnamefont{Hautoj\"arvi}},
  \bibinfo{author}{\bibfnamefont{C.}~\bibnamefont{Corbel}},
  \bibinfo{author}{\bibfnamefont{L.~N.} \bibnamefont{Pfeiffer}},
  \bibnamefont{and} \bibinfo{author}{\bibfnamefont{P.~H.}
  \bibnamefont{Citrin}}, \bibinfo{journal}{Phys.\ Rev. B}
  \textbf{\bibinfo{volume}{54}}, \bibinfo{pages}{R11050}
  (\bibinfo{year}{1996}).

\bibitem[{\citenamefont{Schaefer et~al.}(1986)\citenamefont{Schaefer, Stuck,
  Banhart, and Bauer}}]{schaefer86}
\bibinfo{author}{\bibfnamefont{H.~E.} \bibnamefont{Schaefer}},
  \bibinfo{author}{\bibfnamefont{W.}~\bibnamefont{Stuck}},
  \bibinfo{author}{\bibfnamefont{F.}~\bibnamefont{Banhart}}, \bibnamefont{and}
  \bibinfo{author}{\bibfnamefont{W.}~\bibnamefont{Bauer}}, in
  \emph{\bibinfo{booktitle}{Proceedings of the 8th International Conference on
  Vacancies and Interstitials in Metals and Alloys}}, edited by
  \bibinfo{editor}{\bibfnamefont{C.}~\bibnamefont{Ambromeit}} \bibnamefont{and}
  \bibinfo{editor}{\bibfnamefont{H.}~\bibnamefont{Wollenberger}}
  (\bibinfo{publisher}{Trans Tech}, \bibinfo{address}{Aedermannsdorf},
  \bibinfo{year}{1986}).

\bibitem[{\citenamefont{Hautoj\"arvi et~al.}(1982)\citenamefont{Hautoj\"arvi,
  Johansson, Vehanen, Yli-Kauppila, Hillairet, and
  Tzan\'etakis}}]{Hautojarvi82}
\bibinfo{author}{\bibfnamefont{P.}~\bibnamefont{Hautoj\"arvi}},
  \bibinfo{author}{\bibfnamefont{J.}~\bibnamefont{Johansson}},
  \bibinfo{author}{\bibfnamefont{A.}~\bibnamefont{Vehanen}},
  \bibinfo{author}{\bibfnamefont{J.}~\bibnamefont{Yli-Kauppila}},
  \bibinfo{author}{\bibfnamefont{J.}~\bibnamefont{Hillairet}},
  \bibnamefont{and}
  \bibinfo{author}{\bibfnamefont{P.}~\bibnamefont{Tzan\'etakis}},
  \bibinfo{journal}{Appl. Phys. A} \textbf{\bibinfo{volume}{27}},
  \bibinfo{pages}{49} (\bibinfo{year}{1982}).

\bibitem[{\citenamefont{Manninen and Nieminen}(1981)}]{Manninen81}
\bibinfo{author}{\bibfnamefont{M.}~\bibnamefont{Manninen}} \bibnamefont{and}
  \bibinfo{author}{\bibfnamefont{R.~M.} \bibnamefont{Nieminen}},
  \bibinfo{journal}{Appl. Phys A} \textbf{\bibinfo{volume}{26}},
  \bibinfo{pages}{93} (\bibinfo{year}{1981}).

\bibitem[{\citenamefont{Kuitunen et~al.}(2007)\citenamefont{Kuitunen, Saarinen,
  and Tuomisto}}]{Kuitunen07}
\bibinfo{author}{\bibfnamefont{K.}~\bibnamefont{Kuitunen}},
  \bibinfo{author}{\bibfnamefont{K.}~\bibnamefont{Saarinen}}, \bibnamefont{and}
  \bibinfo{author}{\bibfnamefont{F.}~\bibnamefont{Tuomisto}},
  \bibinfo{journal}{Phys. Rev. B} \textbf{\bibinfo{volume}{75}},
  \bibinfo{pages}{045210} (\bibinfo{year}{2007}).

\bibitem[{\citenamefont{Polity et~al.}(1998)\citenamefont{Polity, B\"orner,
  Huth, Eichler, and Krause-Rehberg}}]{Polity98}
\bibinfo{author}{\bibfnamefont{A.}~\bibnamefont{Polity}},
  \bibinfo{author}{\bibfnamefont{F.}~\bibnamefont{B\"orner}},
  \bibinfo{author}{\bibfnamefont{S.}~\bibnamefont{Huth}},
  \bibinfo{author}{\bibfnamefont{S.}~\bibnamefont{Eichler}}, \bibnamefont{and}
  \bibinfo{author}{\bibfnamefont{R.}~\bibnamefont{Krause-Rehberg}},
  \bibinfo{journal}{Phys.\ Rev.\ B} \textbf{\bibinfo{volume}{58}},
  \bibinfo{pages}{10363} (\bibinfo{year}{1998}).

\bibitem[{\citenamefont{Pesola et~al.}(1999)\citenamefont{Pesola, von Boehm,
  Mattila, and Nieminen}}]{Pesola99}
\bibinfo{author}{\bibfnamefont{M.}~\bibnamefont{Pesola}},
  \bibinfo{author}{\bibfnamefont{J.}~\bibnamefont{von Boehm}},
  \bibinfo{author}{\bibfnamefont{T.}~\bibnamefont{Mattila}}, \bibnamefont{and}
  \bibinfo{author}{\bibfnamefont{R.~M.} \bibnamefont{Nieminen}},
  \bibinfo{journal}{Phys. Rev. B} \textbf{\bibinfo{volume}{60}},
  \bibinfo{pages}{11449} (\bibinfo{year}{1999}).

\bibitem[{\citenamefont{Polity and Rudolf}(1999)}]{Polity99}
\bibinfo{author}{\bibfnamefont{A.}~\bibnamefont{Polity}} \bibnamefont{and}
  \bibinfo{author}{\bibfnamefont{F.}~\bibnamefont{Rudolf}},
  \bibinfo{journal}{Phys. Rev. B} \textbf{\bibinfo{volume}{59}},
  \bibinfo{pages}{10025} (\bibinfo{year}{1999}).

\bibitem[{\citenamefont{W\"urschum et~al.}(1989)\citenamefont{W\"urschum,
  Bauer, Maier, Seeger, and Schaefer}}]{Wurschum89}
\bibinfo{author}{\bibfnamefont{R.}~\bibnamefont{W\"urschum}},
  \bibinfo{author}{\bibfnamefont{W.}~\bibnamefont{Bauer}},
  \bibinfo{author}{\bibfnamefont{K.}~\bibnamefont{Maier}},
  \bibinfo{author}{\bibfnamefont{A.}~\bibnamefont{Seeger}}, \bibnamefont{and}
  \bibinfo{author}{\bibfnamefont{H.-E.} \bibnamefont{Schaefer}},
  \bibinfo{journal}{J. Phys.: Condens. Matter} \textbf{\bibinfo{volume}{1}},
  \bibinfo{pages}{33} (\bibinfo{year}{1989}).

\bibitem[{\citenamefont{Moser et~al.}(1985)\citenamefont{Moser, Pautrat,
  Corbel, and Hautoj\"arvi}}]{Moser85}
\bibinfo{author}{\bibfnamefont{P.}~\bibnamefont{Moser}},
  \bibinfo{author}{\bibfnamefont{J.~L.} \bibnamefont{Pautrat}},
  \bibinfo{author}{\bibfnamefont{C.}~\bibnamefont{Corbel}}, \bibnamefont{and}
  \bibinfo{author}{\bibfnamefont{P.}~\bibnamefont{Hautoj\"arvi}}, in
  \emph{\bibinfo{booktitle}{Positron annihilation}}, edited by
  \bibinfo{editor}{\bibfnamefont{P.~C.} \bibnamefont{Jain}},
  \bibinfo{editor}{\bibfnamefont{R.~M.} \bibnamefont{Singru}},
  \bibnamefont{and}
  \bibinfo{editor}{\bibfnamefont{P.}~\bibnamefont{Gopinathan}}
  (\bibinfo{publisher}{World Scientific}, \bibinfo{address}{Singapore},
  \bibinfo{year}{1985}).

\bibitem[{\citenamefont{Puska}(1989)}]{Puska89b}
\bibinfo{author}{\bibfnamefont{M.~J.} \bibnamefont{Puska}},
  \bibinfo{journal}{J. Phys.: Condens. Matter} \textbf{\bibinfo{volume}{1}},
  \bibinfo{pages}{7347} (\bibinfo{year}{1989}).

\bibitem[{\citenamefont{Xu and Lindefelt}(1990)}]{Xu90}
\bibinfo{author}{\bibfnamefont{H.}~\bibnamefont{Xu}} \bibnamefont{and}
  \bibinfo{author}{\bibfnamefont{U.}~\bibnamefont{Lindefelt}},
  \bibinfo{journal}{Phys. Rev. B} \textbf{\bibinfo{volume}{41}},
  \bibinfo{pages}{5979} (\bibinfo{year}{1990}).

\bibitem[{\citenamefont{El-Mellouhi and Mousseau}(2005)}]{El-Mellouhi05}
\bibinfo{author}{\bibfnamefont{F.}~\bibnamefont{El-Mellouhi}} \bibnamefont{and}
  \bibinfo{author}{\bibfnamefont{N.}~\bibnamefont{Mousseau}},
  \bibinfo{journal}{Phys.\ Rev.\ B} \textbf{\bibinfo{volume}{71}},
  \bibinfo{pages}{125207} (\bibinfo{year}{2005}).

\bibitem[{\citenamefont{Hautakangas et~al.}(2003)\citenamefont{Hautakangas,
  Oila, Alatalo, Saarinen, Liszkay, Seghier, and Gislason}}]{Hautakangas03}
\bibinfo{author}{\bibfnamefont{S.}~\bibnamefont{Hautakangas}},
  \bibinfo{author}{\bibfnamefont{J.}~\bibnamefont{Oila}},
  \bibinfo{author}{\bibfnamefont{M.}~\bibnamefont{Alatalo}},
  \bibinfo{author}{\bibfnamefont{K.}~\bibnamefont{Saarinen}},
  \bibinfo{author}{\bibfnamefont{L.}~\bibnamefont{Liszkay}},
  \bibinfo{author}{\bibfnamefont{D.}~\bibnamefont{Seghier}}, \bibnamefont{and}
  \bibinfo{author}{\bibfnamefont{H.~P.} \bibnamefont{Gislason}},
  \bibinfo{journal}{Phys. Rev. Lett.} \textbf{\bibinfo{volume}{90}},
  \bibinfo{pages}{137402} (\bibinfo{year}{2003}).

\end{thebibliography}

\end{document}